\documentstyle[twoside,fleqn,espcrc2]{article}

\input epsf.sty
\newdimen\psfigsize

\def\psfigure#1 #2 #3 #4 #5{
    \begin{figure}[t]
    \vbox{
    \null\vskip-0.1in\hskip#2\epsfxsize=#1 \epsfbox[50 50 266 201]{#4}
    \vskip -0.35in
    \caption {#5 \label{#3}}
    \vskip -0.22in
    \vskip 0.0truein plus0.2truein}
    \end{figure}
}

\def\NP{{Nucl.\ Phys.\ }}
\def\PL{{Phys.\ Lett.\ }}
\def\PR{{Phys.\ Rev.\ }}
\def\PRL{{Phys.\ Rev.\ Lett.\ }}
\newcommand{\Tr} {\mbox{Tr}}

\newcommand{\AmS}{{\protect\the\textfont2
  A\kern-.1667em\lower.5ex\hbox{M}\kern-.125emS}}

\title{
\vskip -104pt
 \mbox{} \hfill BI-TP 97/37\\
 \mbox{} \hfill September 1997\\
\vskip  70pt
Screening Masses and Improvement in Pure SU(2) Lattice Gauge
  Theory at High Temperatures}

\author{
  U. M. Heller\address{SCRI, Florida State University, Tallahassee,
    FL 32306-4130, USA},
  F. Karsch\address{Fakult\"at f\"ur Physik, Universit\"at Bielefeld,
    P.O. Box 100131, D-33501 Bielefeld, Germany}
  and
  J. Rank\thanks{Talk presented at the Lattice 97 conference by J.R.}
  $^{\mbox{\scriptsize a,b}}$
  }
       
\begin{document}

\begin{abstract}
From the long-distance behaviour of gluon and Polyakov loop correlation
functions we extract masses resp.\ energies in the electric and magnetic
sectors. We discuss their dependence on the temperature and on the momentum
as well as the relevance of an improvement of the lattice discretization of
the action.
\end{abstract}

\maketitle

\section{Introduction}
The high temperature plasma phase of QCD is characterized by the occurrence of
chromo-electric and -magnetic screening masses ($m_e$ and $m_m$) which control
the infrared behaviour of the theory \cite{Lin80}. It has been known for a long
time that the lowest order perturbative result for the electric mass in pure
gauge theory is $m_{e,0}(T) = \sqrt{N_c/3} \, g(T) \, T$. This is sufficient to
cure infrared divergences of ${\cal O}(gT)$. The magnetic mass is known to
be entirely of non-perturbative origin, as all orders in perturbation theory
would contribute equally. However, a dependence of the form
$m_m \sim {\cal O}(g^2T)$ is widely believed and would cure the higher order
infrared divergences of ${\cal O}(g^2T)$. If one finds indeed $m_m \ne 0$, than
$m_m$ contributes in next-to-leading order perturbation theory to
$m_e$ \cite{Re9394},
\begin{eqnarray}
m^2_e(T) \mbox{\hspace*{-1.5ex}} & = & \mbox{\hspace*{-1.5ex}}
m^2_{e,0} \left( 1 + \right. \nonumber \\
& & \mbox{\hspace*{-8ex}}\left. \frac{\sqrt{6}}{2 \pi} \, g(T)
\frac{m_e}{m_{e,0}} \left[ \log \frac{2 \, m_e}{m_m} - \frac{1}{2} \right] +
{\cal O}(g^2) \right) \; .
\label{rebscreen}
\end{eqnarray}
We note that if $m_m \sim {\cal O} (g^2T)$
the next-to-leading order correction to $m_e$ is ${\cal O} (g\ln g)$. 

The above discussion shows that non perturbative methods are needed to obtain
results not only for the magnetic but also for the electric mass. In the
following we will present results for $m_e$ and $m_m$ that we calculated in
SU(2) lattice gauge theory, using both the Wilson action and a tree-level
Symanzik improved action with a planar 6-link loop. Simulations have been
performed on lattices of sizes
$32^3 \!\times\! 4$ and $32^2 \!\times\! 64 \!\times\! 8$ at
temperatures above the critical temperature of the deconfinement phase
transition from $T \!=\! 2 \, T_c$ up to very high temperatures,
$T \!\simeq\! 10^4 T_c$, in order to get in contact with perturbative
predictions for $m_e$.

\section{Screening Masses from Gluon Correlation Functions}
It was shown in \cite{KoKuRe9091} that the pole mass definition of the
screening masses,
\begin{equation}
m_\mu^2 = \Pi_{\mu \mu} (p_0=0, \vec{p}\,^2 = -m_\mu^2) \quad ,
\label{scrmass}
\end{equation}
is gauge invariant although the gluon polarization tensor $\Pi_{\mu \mu}$
itself is a gauge dependent quantity. These pole masses can be obtained from
the long distance behaviour of momentum dependent gluon correlation functions
in the static sector ($p_0 = 0$),
\begin{equation}
\tilde{G}_\mu(p_\bot,x_3)  = \left\langle \Tr \; \tilde{A}_\mu (p_\bot,x_3) 
\tilde{A}_\mu^\dagger (p_\bot,0) \right\rangle
\label{gtilde}
\end{equation}
with the momentum dependent gauge fields
\begin{equation}
\tilde{A}_\mu (p_\bot,x_3)  \,=\!
\sum_{x_0, x_\bot} e^{i \, x_\bot p_\bot} A_\mu(x_0,x_\bot,x_3) \quad .
\label{atilde}
\end{equation}
We define $p_\bot \!\equiv\! (p_1,p_2), x_\bot \!\equiv\! (x_1,x_2)$.

As $\tilde{G}_\mu$ is gauge dependent one has to work in a fixed gauge, which
is in our case the Landau gauge. Details on the gauge fixing algorithm can be
found in \cite{HeKaRa95}. The relation between $\tilde{G}_\mu$ and $m_\mu$ and
techniques how to extract $m_\mu$ most reliable from a lattice simulation are
discussed in \cite{HeKaRa97}.

In Fig.\ \ref{me_gluon}
\psfigure 2.95in 0.0in {me_gluon} {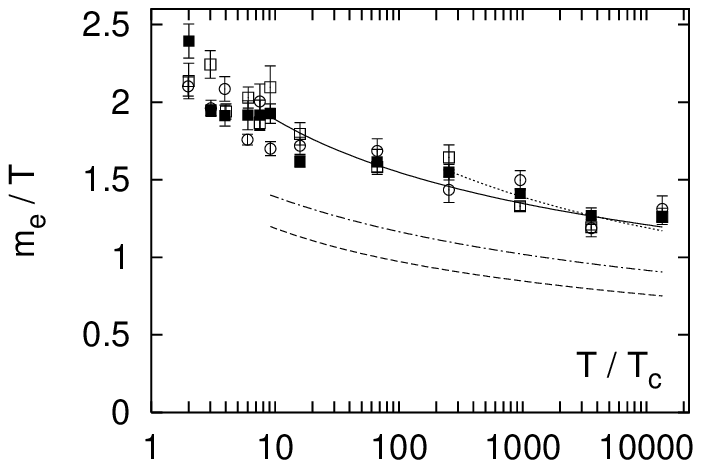} {Electric screening mass in
units of the temperature vs.\ $T/T_c$; see text for details.}
we show the electric screening mass in units of the temperature vs.\ $T/T_c$.
The Wilson action data are symbolized by filled ($N_\tau \!=\! 8$) resp.\ open
squares ($N_\tau \!=\! 4$), the data based on the tree-level Symanzik improved
action with $N_\tau \!=\! 4$ by open circles.

Within statistical errors, all these data are compatible. Especially the fact
that an improvement of the action does not shift the data in any direction is
a first hint that the electric mass indeed is entirely dominated by low
momentum modes. A comparison of the data with the lowest order perturbative
prediction (the dashed line) shows that this is not a good description
although the functional  dependence on the temperature does seem to describe
the data well. For $T \ge 9 \, T_c$ we performed a one par\-ameter fit to our
data with the ansatz $m^2_e(T) = A_{\mbox{\scriptsize fit}} \; g^2(T) \, T^2$,
using the two-loop $\beta$-function for the running coupling. The result
$A_{\mbox{\scriptsize fit}} = 1.69(2)$ (with $\chi^2 / \mbox{dof} = 4.51$),
which is shown as a solid line in Fig.~\ref{me_gluon}, exceeds the perturbative
value $2/3$ by a factor of more than 2.5.

To test the next-to-leading order result (\ref{rebscreen}) we also
calculated the magnetic mass $m_m$. We did this for the Wilson action
with $N_\tau \!=\! 8$. Our results for the ratio $m^2_e/m^2_m$ are shown in
Fig.~\ref{g2}.
\psfigure 2.95in 0.0in {g2} {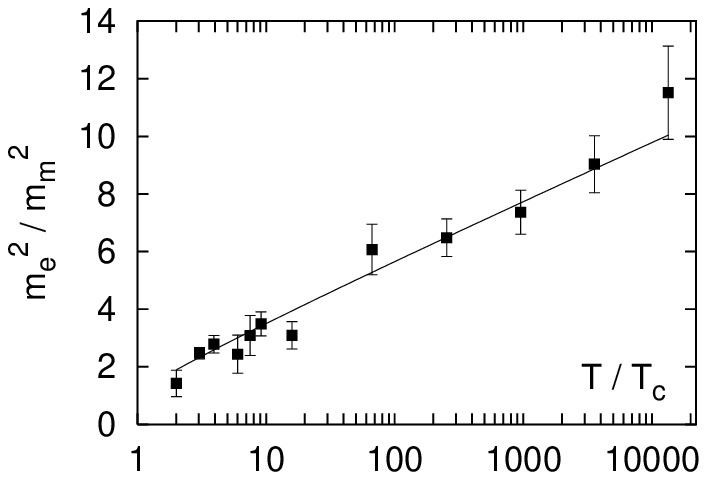} {Squared ratio of electric and magnetic
screening masses vs.\ $T/T_c$ on a $32^2 \times 64 \times 8$ lattice using
the Wilson action.}
They suggest that the ratio runs with the temperature as expected,
$m^2_e/m^2_m \sim g^{-2}(T)$. A fit over the entire temperature range yields
$m^2_e/m^2_m = 7.43(27) g^{-2}(T)$. The quality of the fit becomes obvious
from the small value of $\chi^2 / \mbox{dof} = 1.39$.

Based on this result we fitted the magnetic mass itself, using the ansatz
$m_m(T) \sim g^2(T) \, T$. In the temperature range above $3 T_c$ we obtain
$m_m(T) = 0.457(6) \; g^2(T) \, T$ with $\chi^2 / \mbox{dof} = 1.50$. This is
in good agreement with our result obtained in \cite{HeKaRa95} for
$T < 20 \, T_c$.

With the numerical result for $m_e/m_m$ we are now able to compute $m_e$ in
next-to-leading order, using Eq.~(\ref{rebscreen}). The result is shown by
the dashed-dotted line in Fig.~\ref{me_gluon}. It lies about 20\% above the
tree-level result. However, it is still too small to describe the data.
Therefore we have performed another fit of the electric mass, using the
ansatz
\begin{eqnarray}
\left( \frac{m_e(T)}{T} \right)^2 \mbox{\hspace*{-1.5ex}}
& = & \mbox{\hspace*{-1.5ex}} \frac{2}{3} \, g^2(T) \,
\left( 1 + \frac{\sqrt{6}}{2 \pi} \, g(T) \cdot \right. \nonumber \\
& & \mbox{\hspace*{-10ex}} \left. 
\left[ \log \frac{2 \, m_e}{m_m} - \frac{1}{2} \right] \right)
+ B_{\mbox{\scriptsize fit}} \; g^4(T) \quad .
\label{rebhan_g4_fit}
\end{eqnarray}
As the $g^4$ correction term leads to a temperature dependence which is too
strong within the entire $T$-interval we have restricted the fit to very
high temperatures, $T \ge 250 \, T_c$. The fit gives
$B_{\mbox{\scriptsize fit}} = 0.744(28)$ with $\chi^2 / \mbox{dof} = 4.55$
(dotted line in Fig.~\ref{me_gluon}).

So far we have only discussed the screening masses which had been extracted
from zero momentum correlation functions. In addition, we also measured
gluon correlation functions at finite momenta $p_1 a = 2 \pi \, k_1 / N_1$ with
$k_1 = 1,2$. In order to analyze modifications of the free particle dispersion
relation, which arise from interactions in a thermal medium, we introduce a
parameter $\kappa$ in the dispersion relation:
\begin{equation}
\sinh^2 \frac{a E_e(p_1)}{2} = \sinh^2 \frac{a m_e}{2} +
\kappa \, \sin^2 \frac{a p_1}{2} \quad .
\label{eq:disprel_mod2}
\end{equation}
For $m_e$ we use the results from the $\vec{p}=0$ measurements.
For $T \to \infty$ one expects that the dispersion
relation approaches that of a free particle, i.e.\
$\kappa \to 1$. In the temperature interval analysed by
us, however, we do not observe any statistically
significant increase in $\kappa$. We therefore only quote
a value averaged over the 
temperature interval $T\ge 9 \, T_c$. We obtain  $\kappa = 0.37(10)$ for
$k_1 = 1$ and $\kappa = 0.65(3)$ for $k_1 = 2$.

\section{Polyakov Loop Correlation Functions}
For temperatures above $T_c$ the relation between the electric (or Debye)
screening mass and the colour singlet potential $V_1$ is in lowest order
perturbation theory given by \cite{McSv81}
\begin{equation}
V_1 (R,T) =  - g^2 \, \frac{N^2_c - 1}{8 \pi N_c} \cdot
\frac{e^{- m_e(T) R}}{R}
\label{deconfinement_potential}
\end{equation}
which is again valid only at large distances. On the lattice one can extract
$V_1$ by measuring Polyakov loop correlation functions 
\begin{equation}
e^{- V_1(R,T)/T} = 2 \, \frac{ \langle \Tr \,( L(\vec{R}) \,
L^{\dagger}(\vec{0})) \rangle }{ \langle | L | \rangle^2} \quad .
\label{v1_correlation}
\end{equation}
In addition, we are using not only $V_1$ to extract $m_e$ but also
$V_{1,\mbox{\scriptsize sum}}$ which is based on Polyakov loops that are
averaged over hyperplanes \cite{HeKaRa97}.

The numerical values for $m_e$ using this methods agree within errors with
the data obtained from the gluon correlation functions at zero momentum. The
values of the fits of $m_e/T$ using the two fit ans\"atze described in the last
section are listed in \cite{HeKaRa97}.

To check whether or not an improvement of the action weakens the violation of
the rotational symmetry caused by the lattice discretization we measured
$V_1 / T$ both along an axis, labeled with $(1,0,0)$, and along three
different off-axis directions, $(1,1,0)$, $(1,1,1)$, and $(2,1,0)$.

For each action we computed the fit of the $(1,0,0)$ data from which we
calculated the $\chi^2$ deviation of the off-axis data. For all 12
temperatures we observed much larger deviations in the case of Wilson
action than for the tree-level Symanzik improved action. In addition, the
parameters of the fits of the $(1,0,0)$ data themselves have bigger errors and
furthermore the fits have larger $\chi^2$ and lower goodness in the case of
Wilson action than for the improved action.

\section{Summary}

We have studied electric and magnetic screening masses obtained from Polyakov
loop and gluon correlation functions in the high temperature deconfined phase
of SU(2) lattice gauge theory, using both the standard Wilson action and a
tree-level Symanzik improved action.

For $m_m$ we find the expected ${\cal O}(g^2T)$ behaviour,
$m_m(T) = 0.457(6) \, g^2(T) \, T$. We also find
$(m_e / m_m)^2 \!\sim\! g^{-2}$,
which suggest that the temperature dependence of $m_e$ is well described by
$m_e/T \!\sim\! g(T)$ as expected by lowest order PT. On a
quantitative level we do, however, find large deviation. Our result
$m_e(T) = \sqrt{1.69(2)} \, g(T) \, T$, deviates strongly from the
lowest order PT prediction, and it can only be insufficiently cured by the
next-to-leading order term. 

The improvement of the action does not show, within statistical errors, any
significant modification in the behaviour of the screening masses, although we
find that the violation of the rotational symmetry of the singlet potential,
which also is used to extract $m_e$, is weakened.


\begin{thebibliography}{9}

\bibitem{Lin80}
A.D.~Linde, \PL {\bf B96} (1980) 289.

\bibitem{Re9394}
A.K.~Rebhan, \PR {\bf D48} (1993) R3967 and \NP {\bf B430} (1994) 319.

\bibitem{KoKuRe9091}
R.~Kobes, G.~Kunstatter, A.~Rebhan, \PRL {\bf 64} (1990) 2992 and
\NP {\bf B355} (1991) 1.

\bibitem{HeKaRa95}
U.M.~Heller, F.~Karsch and J.~Rank, \PL {\bf B355} (1995) 511.

\bibitem{HeKaRa97}
U.M.~Heller, F.~Karsch and J.~Rank, BI-TP 97/36, September 1997.

\bibitem{McSv81}
L.D.~McLerran and B.~Svetitsky, \PR {\bf D24} (1981) 450.

\end{thebibliography}
\end{document}